\documentclass[11pt]{article}

\usepackage{times}

\usepackage{amsmath}
\usepackage{amsfonts}
\usepackage{amssymb}
\usepackage{graphicx}
\usepackage[longnamesfirst]{natbib}
\setcitestyle{aysep={},notesep=:}
\usepackage{endnotes}
\usepackage[hidelinks,bookmarks=false]{hyperref}

\begin{document}

\renewcommand\footnotesize{\normalsize}

\title{Frame-Induced Group Polarization in Small Discussion Networks}

\author
{Michael Gabbay,${}^{1\ast}$ Zane Kelly,${}^{1}$ Justin Reedy,${}^{2}$ John Gastil${}^{3}$ \\ \\
\normalsize{${}^{1}$Applied Physics Laboratory,} \\
\normalsize{University of Washington, Seattle, WA 98105, USA}\\
\normalsize{${}^{2}$Dept. of Communication and Center for Risk \& Crisis Management,} \\
\normalsize{University of Oklahoma, Norman, OK 73019, USA}\\
\normalsize{${}^{3}$Depts. of Communication Arts \& Sciences and Political Science,} \\
\normalsize{Pennsylvania State University, University Park, PA 16802, USA}\\
\\
\normalsize{$^\ast$To whom correspondence should be addressed; E-mail:  gabbay@uw.edu.}
}

\date{}

%
%
%
%
%

\maketitle

\shortcites{McGTurHog1992,DefAmbWei2002,XieSreKor2011,WaaVerCha2015,DavHulAu1997,DavZarHul1997,Mouetal2013}

\begin{abstract}
We present a novel explanation for the group polarization effect whereby discussion among like-minded individuals induces shifts toward the extreme. Our theory distinguishes between a quantitative policy under debate and the discussion's rhetorical frame, such as the likelihood of an outcome. If policy and frame position are mathematically related so that frame position increases more slowly as the policy becomes more extreme, majority formation at the extreme is favored, thereby shifting consensus formation toward the extreme. Additionally, use of a heuristic frame can shift the frame reference point away from the policy reference, yielding differential polarization on opposing policy sides. We present a mathematical model that predicts consensus policy given group member initial preferences and network structure. Our online group discussion experiment manipulated policy side, disagreement level, and network structure. The results, which challenge existing polarization theory, are in qualitative and quantitative accord with our theory and model.
\end{abstract}

\vspace{\baselineskip}

\normalsize{\noindent This is the accepted version of the manuscript. The published version is: \\ Gabbay, Michael, Zane Kelly, Justin Reedy, and John Gastil. 2018. ``Frame-Induced Group Polarization in Small Discussion Networks.'' \emph{Social Psychology Quarterly} 81 (3):248--271.  \url{https://doi.org/10.1177/0190272518778784}}

\newpage

Experiments on small groups deciding on acceptable risk levels have shown that discussion can shift group members toward more risky choices. This ``risky shift'' effect was the first demonstration of the broader phenomenon of group polarization in which discussion among members who share the same initial inclinations induces more extreme decisions or opinions. This constitutes ``polarization'' in the sense of shifting further toward one pole rather than divergence toward opposite poles. Hundreds of experiments have shown, for example, that groups also exhibit systematic shifts in the direction of more caution, greater hostility toward an out-group, more lenient punishments, and more extreme political attitudes \citep{MyersLamm1976,Myers1982,Isenberg1986,Brown1986}. The basic precondition for the group polarization effect is that all group members have initial positions on the same side of the neutral reference point of the issue discussed. Group polarization is then said to occur if discussion causes the mean position of the group to shift further away from the reference point as compared with the pre-discussion mean.

In this paper, we present theoretical arguments, mathematical modeling, and experimental results that address how disagreement reduction processes, the reference point, and network structure impact group polarization, effects not well understood in existing research. We propose a new theoretical explanation of group polarization that, while complementary to the standard informational and normative influence mechanisms, integrates majority influence and consensus pressure into its mechanism and reckons more explicitly with the reference point.

We specifically consider deliberations on quantitative policies, such as wagers, investments, interest rates, or military operation sizes (although as we remark in the Discussion, our theory should be applicable  more generally  to attitudes).  The aspect of the policy upon which deliberations focus --- what we call the \emph{rhetorical frame} --- is crucial because group member agreement is driven by proximity along the rhetorical frame rather than the policy. The rhetorical frame will typically correspond to the dominant source of disagreement within the group due, for instance, to uncertainty as to the likelihood of an outcome or the value placed on it. For example, a group of military staff officers discussing how many troops to assign to a planned attack (the policy) may agree on the value of the objective but argue over their different assessments as to how robustly the enemy will defend it (the rhetorical frame). For its simplicity and relevance to our experiment, we concentrate on the case when the rhetorical frame is the subjective probability associated with a binary outcome.

The assumption that groups primarily discuss a dominant rhetorical frame yields two distinct effects for group polarization, one on majority formation and the other on the operative reference point. The first effect, involving \emph{distribution reshaping}, lies at the core of our proposed mechanism for group polarization. As an initial illustration, consider a scenario in which three investors discuss a policy of how much to invest in a startup company. Prior to discussion, they have initial investment preferences of \$4M, \$5M, and \$6M. If the key uncertainty is whether the company will successfully bring its product to market, the rhetorical frame should be the subjective probability of that outcome. The three investors are taken to have initial respective subjective probabilities of company success of .6, .7, and .75. These probabilities, being above .5, imply that the investors are like-minded in the sense that they all believe the company more likely to be successful than not. Note that the distribution of group member positions along the rhetorical frame (their ``rhetorical positions'') has a different shape than the policy distribution: the latter is symmetric, not favoring either side of the mean of \$5M, while the rhetorical position distribution is asymmetric, having two values above the mean (.683) and one below. Furthermore, the rhetorical distance between the members of the low investment pair $(\$4M,\$5M)$ is .1, twice the distance of .05 in the high investment pair $(\$5M,\$6M)$. Since agreement is driven by proximity of the rhetorical positions, the higher pair is more likely to form the group majority that drives the consensus policy. Assuming this pair forms a majority around some policy intermediate between them (e.g., \$5.5M), the final consensus investment will be higher than the initial mean, signifying group polarization.

More generally, distribution reshaping occurs because the relative spacing of group member positions along the rhetorical frame need not be the same as along the policy itself and so the distribution of rhetorical positions may be reshaped in comparison with the distribution of policy preferences. In particular, for a concave (downward curvature) functional dependence of rhetorical position on policy, the rhetorical position increases more slowly as the policy becomes more extreme (as does the subjective probability as a function of investment in the above example). Therefore, the rhetorical distance between more extreme members is compressed relative to more moderate ones. This compression facilitates the emergence of a majority at the extreme, which then drives consensus to a policy that is more extreme than the mean of the members' initial policy preferences. As applicable to our experiment, we use the theory of decision making under risk and uncertainty to show that the subjective probability rhetorical frame has the requisite concave dependence upon policy for generating polarization.

The second theorized effect is heuristic substitution of an intuitive frame for a more complex rhetorical frame appropriate to the policy. This \emph{heuristic frame substitution} can result in discussion of a rhetorical frame whose reference point is offset from the policy reference. The rhetorical reference then resides within either the pro or con policy side rather than at the boundary between them. The side split by the heuristic frame reference is subject to weaker distribution reshaping and so polarization is suppressed on that side (but enhanced on the other).

We implement our theory mathematically using a rhetorically-proximate majority (RPM) model. The RPM model seeks to predict group member final policies from their initial ones while accounting for network structure. The rhetorically-proximate majority is the subset of group members that comprises the majority spanning the least rhetorical distance. The consensus policy is then given by a weighted average of the policy preferences within the subset. If the structure of the group's communication network is such that some members have more ties than others, then more central members have greater weights in determining the consensus policy.

\begin{figure}[p]
\begin{center}
\includegraphics[width=0.7\textwidth]{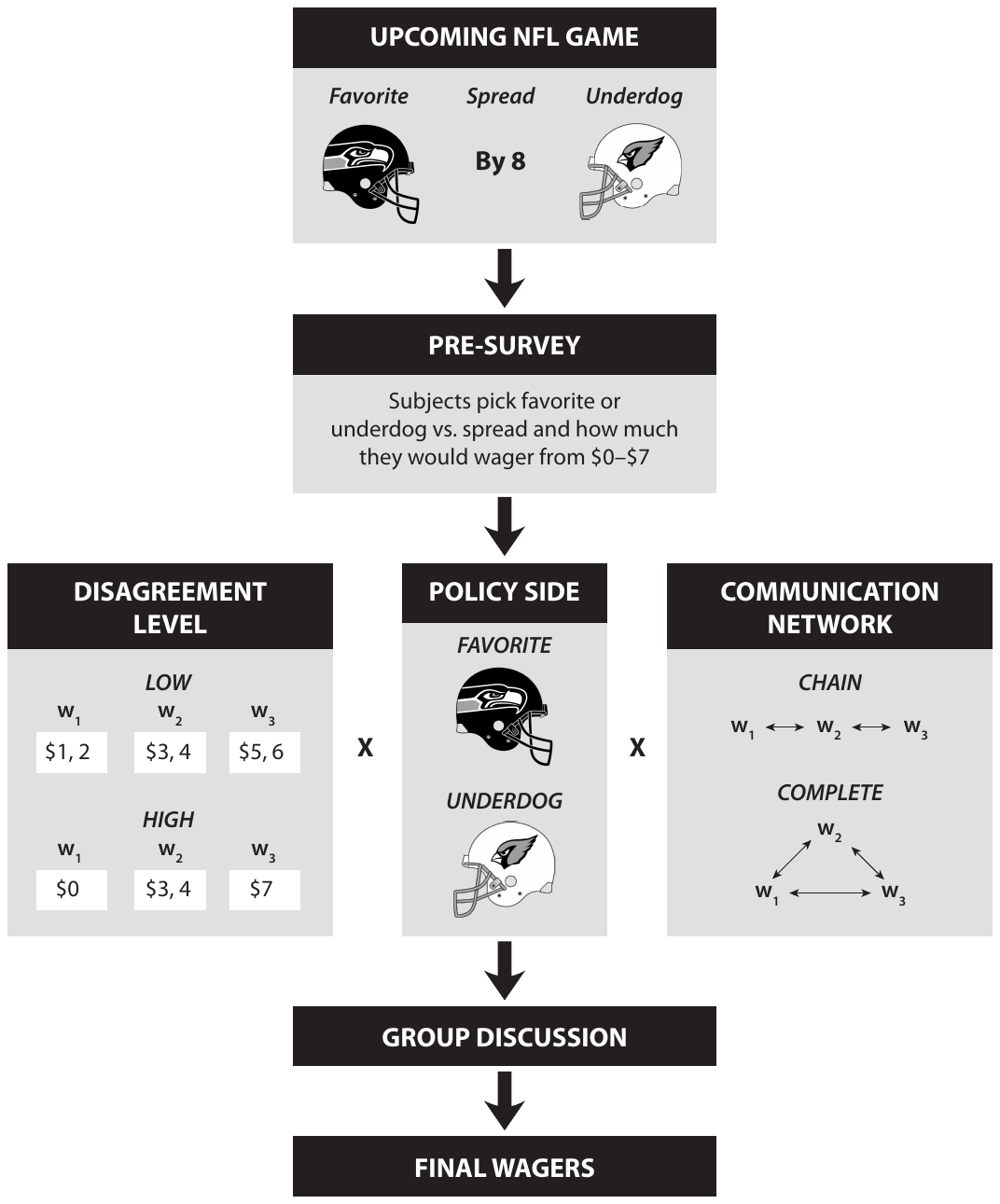}
\caption{Experiment setup. The Seahawks are shown as the favorite by $8$ points over the Cardinals. A bet on the Seahawks is successful if they win by more than 8 points; otherwise, a bet on the Cardinals is successful. Two values shown for a $w_i$ indicates either value is permissible.}\label{fig:setup}
\end{center}
\end{figure}

We conducted an experiment in which triads engaged in online discussion about wagering on National Football League (NFL) games (Figure~\ref{fig:setup}). As is standard practice in football betting, spread betting was employed rather than wagering directly on which team will win the game. In spread betting, the terms ``favorite'' and ``underdog'' are used to refer, respectively, to the likely winner and loser of the game itself. The point spread is the expected margin of victory of the favorite team as set by Las Vegas oddsmakers. A bet on the favorite is successful if its margin of victory exceeds the spread; otherwise a bet on the underdog is successful.\endnote{In actual practice, bets are returned if the victory margin equals the spread.} On the basis of a pre-survey that elicited team choices and wager amounts, discussion groups were constructed with respect to three dichotomous variables: (1) \emph{policy side} of favorite or underdog corresponding to the team chosen as more likely to beat the spread; (2) \emph{disagreement level} of ``high'' or ``low'' between the minimum and maximum wagers in the group; and (3) \emph{network structure} of ``complete'' in which all members could communicate with each other or ``chain'' in which the members with the lowest and highest wagers were the end nodes of the chain and the intermediate-wager member served as the center node. After discussion, each member made their final wager. A group decision was not required but groups arrived at a consensus wager far more often than the alternative outcomes of a two-person majority or three different wagers.

Defining an increase in mean wager due to discussion as a risky shift, statistically significant results were observed in consensus groups for all three variables: (1) a risky shift only on one policy side (favorite), and greater shifts for (2) high disagreement groups (vs.\ low) and (3) complete networks (vs.\ chains). These results are not readily explained by existing group polarization theory. They are, however, in accord with our theory and the RPM model. The first is consistent with heuristic frame substitution, the second with distribution reshaping and rhetorically-proximate majority formation, and the third with the greater centrality and intermediate initial wager of the chain middle node.

Going beyond the testing of qualitative hypotheses, we also test the quantitative predictions of the RPM model. Simulation using actual group initial wagers to predict consensus wagers passes a statistical goodness-of-fit test. Alternative models (the median, policy-based majority influence, and policy averaging weighted by rhetorically-based confidence) do not provide a satisfactory fit.

\section*{Background}

\subsection*{Group Polarization Theory}
In general, the contexts in which group polarization occurs are on the judgmental side of the intellective-judgmental spectrum in which purely intellective tasks (e.g., math problems) have demonstrably correct solutions and purely judgmental tasks are matters of personal taste \citep{LauAda1982}. The seminal risky shift experiments in the 1960s employed ``choice dilemmas'' in which subjects were presented with hypothetical scenarios involving the choice between a risky but higher payoff option versus a safer, lower payoff one \citep{Brown1986}. Later experiments found that discussion among similarly-inclined individuals produced more extreme social and political attitudes (\citealt[223--24]{Brown1986}; \citealt{SchSunHas2010}; \citealt{KeaVanJud2016}). Manipulation of the evidence presented to mock juries produced discussion-induced shifts to softer or harsher sentences in cases where the evidence was weak or strong respectively \citep[129--30]{Myers1982}. Group polarization has been observed when groups were composed randomly as in the choice dilemmas, given alternative stimuli as in mock juries, or pre-screened to compose homogeneous groups as in attitude studies \citep[604-605]{MyersLamm1976}. Although most experiments have required a unanimous decision, group polarization is still robust even when a group decision is not required \citep[611]{MyersLamm1976}.

The two most widely accepted causal accounts of group polarization offer complementary explanations based on information exchange or social norms \citep{Myers1982,Brown1986,Isenberg1986}.\endnote{Less prominent explanations include social identity theory \citep{McGTurHog1992}, social decision schemes \citep{ZubCroWer1992}, and extremist confidence \citep{Kerr1992}.} Persuasive arguments theory hinges on the role of novel arguments, those not already known to all group members. In essence, group members on the same side of an issue will typically possess different arguments buttressing that side; discussion then exposes group members to new information supporting their initial inclination thereby shifting it further in the same direction. Key supporting evidence for this theory is provided by experiments that show groups polarizing on the basis of arguments only, without exposure to others' preferences \citep{Brown1986}. Persuasive arguments theory also invokes the existence of an imbalance in the cultural pool of arguments to explain polarization over groups drawn \emph{randomly} from the population \citep{VinBur1974}. That is distinct, however, from the novel information exchange process, which only depends on the balance of arguments within a given group.

In social comparison theory, a culturally salient norm prejudices group members' positions toward one pole of the issue. For example, a norm favoring risk-taking makes riskier positions more socially desirable than cautious ones. In the most common variant, when group members become aware of each other's positions, those who are below the mean shift their positions in the direction of the normative ideal. The most telling evidence in support of this theory shows that polarization still occurs when preferences are shared with no discussion whatsoever (the ``mere exposure'' effect, \citealt[1142]{Isenberg1986}).

\subsection*{Open Questions}

This section discusses open questions in standard polarization theory (shorthand for both persuasive arguments and social comparison theories) that are theoretically fundamental and limit its predictive ability in natural settings. We also state its predictions (or lack thereof) for our experiment.

The most important question we consider involves how polarization mechanisms play out in conjunction with processes for resolving disagreement. A strong reading of either persuasive arguments theory or social comparison theory implies that a group will shift toward the extreme whenever its members have homogeneous inclinations (for a judgmental issue) and, respectively, possess distinctive supporting information or operate under a biasing norm --- regardless of their initial opinion distribution. Standard polarization theory has never clearly explicated how the informational and normative mechanisms operate relative to the concurrent --- and potentially countervailing --- processes of majority influence and pressure toward consensus. The informational and normative mechanisms only explain why group members at the low end of an attitude scale preferentially shift their positions upwards, but not how and where the group reaches agreement.  If standard theory's implicit assumption that, absent systematic polarization, a group typically converges to the mean of its members' initial opinion distribution were true, then there would be justification for the strong reading. Convergence to the mean, however, has not been borne out experimentally \citep{DavHulAu1997,Friedkin1999,OhtMasNak2002}. Consequently, when faced with a given group with a specific opinion distribution, standard theory cannot predict what effects its distinctive informational and normative mechanisms will have on the outcome; a group with a moderate majority may depolarize, counter to the strong prediction of uniform polarization.\endnote{\citet{MasFlache2013} have developed an agent-based model of persuasive arguments theory, supplemented by argument forgetting, that can be applied to specific group opinion distributions given knowledge of the underlying number of pro and con arguments.} Although its mechanisms transpire at the group level, standard theory has primarily been operationalized with respect to a population of randomly composed homogeneous groups, over which the effects of (preference-based) majority influence could be assumed to wash out. Because disagreement reduction processes are not integrated into its mechanisms, neither the informational or normative influence theories make a clear prediction as to the effect of disagreement level in our experiment.

Another question of concern involves the reference point, intrinsic to the very definition of polarization.  A group is said to be homogeneous whenever all members are on one side of the reference point; polarization then refers to motion away from the reference. While it is convenient in experiments to set the reference point as the midpoint on an attitude scale, standard theory provides little guidance as to reference point selection criteria that could be applied in natural decision contexts such as foreign policy crises. With respect to our experiment, which composed groups homogeneous with respect to the favorite or underdog policy side, standard theory predicts equal polarization for both sides. Persuasive arguments theory's core process of sharing novel information predicts that both favorite and underdog groups should show a risky shift given that group members presumably all possess different information in support of their common team choice. Social comparison theory observed that low stakes decisions tended to induce a norm toward risk taking \citep[214--15]{Brown1986} and so predicts an increase in wagers regardless of team choice.

Standard theory also does not address how group polarization is affected by communication network structure. Only one previous study has experimentally investigated group polarization with respect to different communication network topologies: no effect of topology was reported although the polarization effect itself was largely absent \citep{Friedkin1999}. Regarding our experiment, while one plausible extension of persuasive arguments theory would predict complete networks to exhibit a greater shift than chains due to the freer flow of information, it could be reasonably expected that the long discussion duration allows for all arguments to be shared in both networks and so both should shift equally.

\section*{Theory}

\subsection*{Rhetorical Frame and Function}

In this section, we discuss the rhetorical frame as distinct from the quantitative policy under discussion and we describe the \emph{rhetorical function} that mathematically relates positions along the frame to policy positions. We show that the rhetorical function relating subjective probability (rhetorical frame) to wager (policy) is expected to be concave. In the following section, we then explain the impact of this concavity in reshaping the frame distribution relative to the policy distribution.

A mathematically idealized caricature of the deliberative process could have group members stating only their numerical policy preferences, then updating their own positions in response (as in the mere exposure effect). A more prosaic depiction has group members exchanging persuasive messages involving facts and reasoning relevant to the policy choice. Persuasive arguments theory essentially counts the number of arguments on both sides of the issue. Instead, our approach is to consider the way the discussion is framed. We define the rhetorical frame as the dominant dimension of the policy problem used during discussion.

In its general sense, the term ``rhetoric'' encompasses the substantive content of persuasive speech not merely its stylistic form. The rhetorical frame is a substantive aspect of the issue but there can be freedom in what frame is discussed. The frame should be a significant underlying dimension of disagreement within the group, reflecting either a fundamental uncertainty in the problem or differences in how members value a particular outcome. In cases where there are several comparable dimensions of disagreement, a single rhetorical frame could still arise due to path-dependent group dynamics, such as efforts to steer a debate's focus.

Although persuasive messages over the rhetorical frame may be expressed in purely qualitative language, we assume that rhetorical positions can be quantitatively related, however loosely, to numerical policy preferences. Accordingly, we represent the rhetorical frame using a one-dimensional axis and define the rhetorical function as the mathematical transformation that maps policy preferences to rhetorical positions. This enables us to formulate an alternative idealization to the simple exchange of policy preferences above in which persuasion, and hence agreement, is driven by the proximity of rhetorical positions.

Essential to our account of group polarization is that the rhetorical function be nonlinear so that rhetorical distance is not uniformly proportional to policy distance. In the example in the introduction, the three investors exhibited a nonlinear rhetorical function (a \$1M investment distance is mapped to subjective probability distances of 0.1 and 0.05 for the low and high pairs respectively). A nonlinear rhetorical function will be common in decision making under risk and uncertainty in which member policy preferences are combinations of the subjective probabilities and utilities of outcomes \citep{PleDieWal2015}. For classic expected utility theory, utility is generally a nonlinear function of wealth. Furthermore, in the more modern prospect theory, the weights used to combine outcome utilities are also nonlinear functions of probability (reflecting psychological tendencies to over or under-weight probabilities). This ubiquity of nonlinearity implies that the relationship between the subjective probability and the optimal policy will commonly be nonlinear.

In particular, we consider decisions for events with two mutually exclusive outcomes. For example, an upcoming election could impinge to an imminent foreign investment decision. The election features two different candidates --- one who seeks to nationalize industry, the other a champion of foreign investment. Group members, even if they have identical utility curves, may differ in their estimates of the likely election outcome and hence have favor different amounts of investment. The rhetorical frame in such cases is the subjective probability of an outcome (the election winner) that stands apart from the policy (the investment amount).

We now proceed to show that a concave rhetorical function is expected in a binary gamble with two equal payoff outcomes where the policy is the wager amount and the rhetorical frame is the subjective probability of the outcome considered to be more likely (as accords with homogeneous preferences). Actually, we first discuss the inverse of the rhetorical function that maps subjective probabilities to wagers to conform with the direction of causal influence.\endnote{The rhetorical function goes in the direction of inference: from an observable variable, the explicitly numerical policy, to an unobserved one, the subjective probability.} We show that the wager displays a convex (upward curvature) dependence on the subjective probability so that the wager increases more rapidly as the subjective probability becomes larger. This implies that the rhetorical function itself is concave.

We use the simple mean-variance approximation of expected utility, which is valid for small wagers \citep[11-12]{EecGolSch2005}.\endnote{Online Appendix~A.1 shows more generally that the convex dependence of wager upon subjective probability arises under both expected utility and prospect theories using an exponential utility function.} It states that the expected utility of a gamble rises with the mean expected gain but diminishes with increasing uncertainty, operationalized as the variance around the mean. As the formal treatment below shows, there are two effects as the subjective probability of the outcome deemed more likely grows: the mean gain increases while the variance decreases. These effects reinforce the expected utility of increasing the wager and lead to an accelerating, hence convex, dependence of wager on subjective probability.

Formally, we consider a binary gamble with two mutually exclusive outcomes, $A$ and $B$, which an individual estimates to have respective probabilities $p$ and $q$ ($=1-p$). Assuming $p>1/2$, for a \$1 wager on $A$, the mean gain is equal to $2p-1$ and the variance is given by $4p(1-p)$, as follows from a binomial distribution. The expected utility $U(w)$ of a wager $w$ is then $U(w) = (2p-1)w - 4\alpha p(1-p)w^2$, where $\alpha$ is an individual's risk aversion (a larger $\alpha$ represents greater sensitivity to uncertainty). For a fixed $w$, the variance is greatest for $p=1/2$. Thus, increasing $p$ has the dual effect of increasing the mean gain while decreasing the variance around it, thereby enabling $w$ to increase as an accelerating function of $p$. More precisely, finding the maximum expected utility by setting the derivative of $U$ with respect to $w$ equal to zero yields the optimal wager as a function of probability:
\begin{equation}
w=\frac{2p-1}{8\alpha p(1-p)}.
\label{eq:meanvar}
\end{equation}
The wager is an accelerating function of probability (for $p>1/2$) given that $w \rightarrow \infty$ as $p \rightarrow 1$. For $p<1/2$, outcome $B$ is more likely and one should replace $p$ by $q$ in Equation~\ref{eq:meanvar} to avoid negative wagers. However, if we employ the artifice of assigning negative wagers to bets on $B$, then we can extend Equation~\ref{eq:meanvar} to represent $B$ bets as well. This allows use of a single policy axis and single rhetorical function, with positive and negative wagers corresponding to opposite policy sides.

\begin{figure}[p]
\begin{center}
\includegraphics[width=0.9\textwidth]{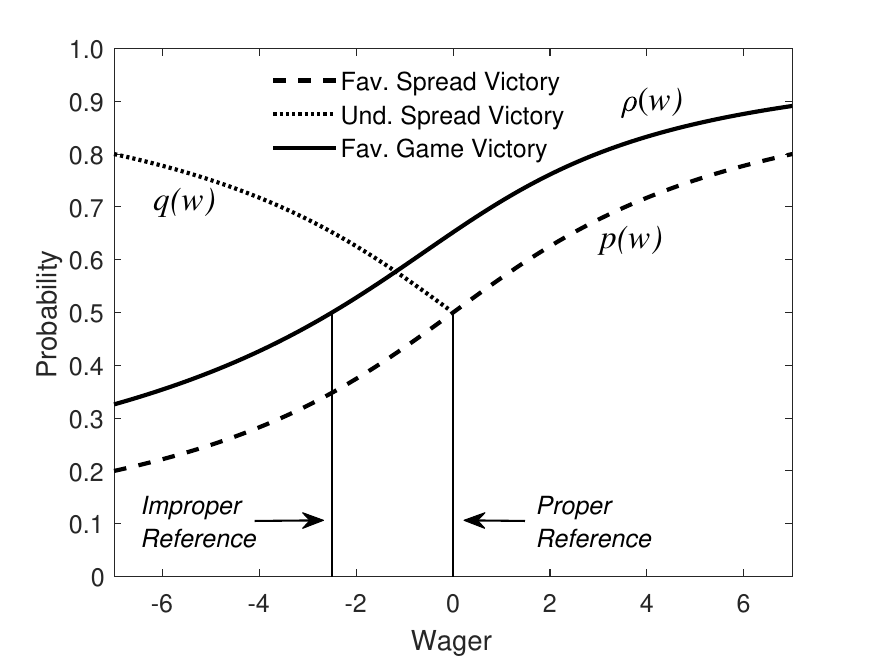}
\caption{Relationship between subjective probabilities and wager amount. Negative wagers correspond to bets on underdogs. Favorite spread victory probability is given by $p(w)$ in Note~\ref{note:pw} with $\alpha=.067$. $q(w)=1-p(w)$. Favorite game victory probability is given by $\rho(w)$ in Note~\ref{note:rhow} with the spread $s_0=5$ and $\sigma=12.8$.}
\label{fig:nflcurves}
\end{center}
\end{figure}

Relevant to our experiment, Equation~\ref{eq:meanvar} is derived in \citet[646]{WooWoo1991} for the specific case of spread betting. As the point spread is set with the intent of equalizing the odds of winning a bet on either team, the payoff is the same regardless of whether one bets on the favorite or underdog \citep{SimmNel2006}.  Taking $A$ above to represent the outcome of the favorite's victory margin exceeding the spread, and $p$ as an individual's subjective probability estimate of that outcome, then Equation~\ref{eq:meanvar} yields the wager, with a positive (negative) wager signifying a bet on the favorite (underdog). Inverting the relationship shows that the subjective probability of a successful favorite bet, $p(w)$, is indeed a concave function of the wager for $w>0$, as plotted in Figure~\ref{fig:nflcurves}.\endnote{\label{note:pw}$p(w) = \frac{1}{2} - \frac{1}{8\alpha w} \pm \frac{1}{2} \sqrt{1 + \frac{1}{16 \alpha^2 w^2}}$ where $+(-)$ corresponds to $w>0(<0)$.} Note that the convexity of $p(w)$ for $w<0$ results from the use of Equation~\ref{eq:meanvar} with the negative underdog wager convention. Directly plotting the probability of a successful underdog bet, $q(w)$, reveals the required concave form as the wager (magnitude) grows. The overall S-shape of $p(w)$ therefore reflects the underlying concavity for both sides of the bet.

\subsection*{Distribution Reshaping} \label{sec:RIA}

If the rhetorical function is nonlinear, then the distribution of rhetorical positions will be reshaped relative to the policy distribution. In particular, a policy distribution that is symmetric around the mean can be skewed into an asymmetric rhetorical distribution favoring either the moderate or extreme ends. Since two policy intervals of the same length but different locations will map onto different length rhetorical position intervals, a nonlinear rhetorical function can cause the relative rhetorical distances between group members to be compressed or expanded as compared with their policy distances. As agreement is forged along the rhetorical frame, distribution reshaping can therefore affect the composition of the majority that emerges during discussion. In the investors triad, for example, the rhetorical distance between the second and third members is compressed relative to that between the first and second members, so we expect the former pair to form a majority more readily than the latter.

\begin{figure}[p]
\centerline{\includegraphics[width=.9\textwidth]{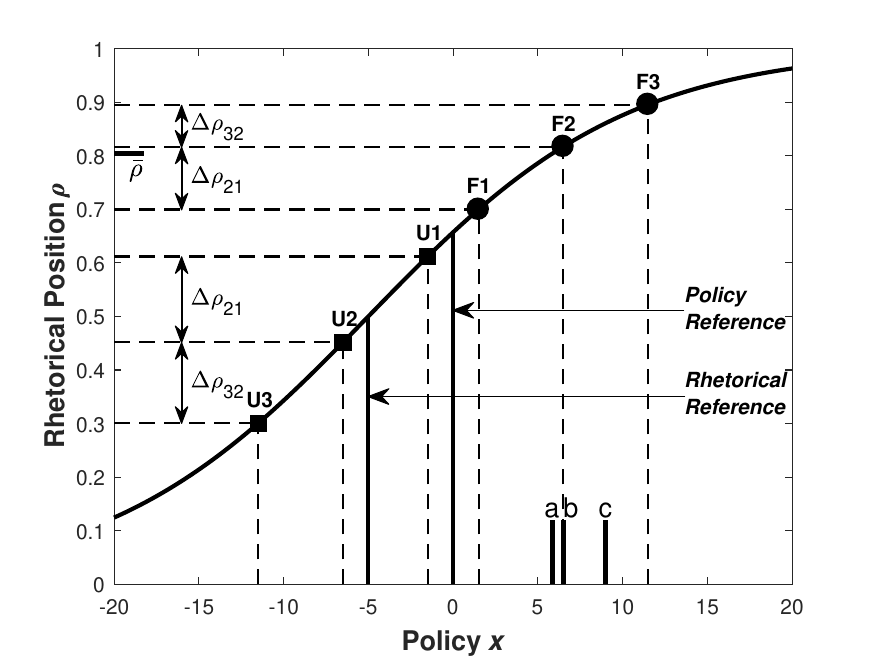}}
\caption{Rhetorical function relating rhetorical position to policy illustrating distribution reshaping and reference point shifting. The $\Delta\rho$'s indicate rhetorical position differences. Short lines at bottom show alternative $F$ group consensus policies: inverse of the rhetorical mean $x(\bar{\rho}) (`a')$; mean policy $\bar{x}=x_2$ (`b'); RPM model policy $\bar{x}_{23}=(x_2+x_3)/2$ (`c').}
\label{fig:riacurve}
\end{figure}

A concave rhetorical function, in which the rhetorical position increases more slowly as the policy increases, will tend to favor majority formation toward the extreme. Figure~\ref{fig:riacurve} illustrates the effect of concavity for a rhetorical function $\rho(x)$ relating rhetorical position $\rho$ to policy $x$. The policy reference at $x=0$ demarcates opposing sides of the policy. Note that the policy and rhetorical references are offset --- a consequence of heuristic frame substitution to be discussed below but not required for the basic distribution reshaping effect. We consider the $F_{1,2,3}$ triad with policy preferences located on the positive policy axis, indicating that they are on the same policy side. The spacing between its members is uniform along the policy axis as the moderate $F_1$ and the extremist $F_3$ are both equidistant from $F_2$. If the consensus process involved a convergence to the policy mean, then the consensus policy should be the policy of the middle member $F_2$. However, because the $F$ group is arrayed on the shoulder of the curve, their rhetorical positions are not uniformly spaced: the intermediate member $F_2$ is rhetorically much closer to the extremist $F_3$ than to the moderate $F_1$. Thus, a symmetric policy distribution is transformed into an asymmetric rhetorical distribution, which sets the stage for the formation of an extreme majority ($F_2$ with $F_3$).

\subsection*{Heuristic Frame Substitution}

The use of heuristics when making judgments has been defined in terms of attribute substitution: a judgment on a target attribute is addressed by using a heuristic attribute more accessible to the mind \citep{KahFre2002}. For example, in assessing the importance of a scientific paper, one might substitute the prestige of the journal in which it appears for the more complex evaluation of its content. This section considers the situation where attribute substitution results in replacing a correct rhetorical frame by a heuristic one.

As indicated in Figure~\ref{fig:riacurve}, we refer to the transformation of the neutral point of the rhetorical frame ($1/2$) onto the policy axis simply as the \emph{rhetorical reference} (the policy reference is always taken to be zero). We define a \emph{proper} rhetorical frame as one whose rhetorical reference coincides with the policy reference so that opposing rhetorical sides fully align with opposing policy sides. In our experiment, the subjective probability of the favorite winning against the spread (being the ``spread victor'') is a proper rhetorical frame. As shown in Figure~\ref{fig:nflcurves}, the neutral point for the subjective probability, $p(w)$, is $1/2$ which indeed transforms to a zero wager, the policy reference. Rhetorical positions in the region $p(w)>1/2$ get mapped to positive wagers and $p(w)<1/2$ map to negative wagers so that the pro and con sides regarding the likelihood of a favorite spread victory align with the favorite and underdog wagers respectively. The spread victor probability therefore results in equivalent distribution reshaping effects for both favorite and underdog sides of the gamble.

Although professional gamblers may be able to directly discuss the spread victor probability, most knowledgeable NFL fans are only prepared to discuss the game victor probability --- directly applying their knowledge of teams' relative strengths and weaknesses to consider which team is most likely to win the game. Therefore, we expect that the game victor probability will be used as a heuristic substitute for the spread victor probability. An empirical basis for expecting this frame substitution in our experiment is provided by the origin of the strong bias toward selecting favorites in NFL spread betting, even though individuals are aware that the spread evens the odds on average, as observed by \citet{SimmNel2006}. They explain this bias via the operation of an ``intuitive confidence'' heuristic in which the initial assessment of a spread bet is guided by an individual's subjective probability concerning who will win the game. Frame substitution then represents the transition of this heuristic from System 1 thinking (fast, intuitive) into the focus of discussion in System 2 thinking (slow, deliberative) \citep{KahFre2002}.

The game victor probability frame is an example of an \emph{improper} rhetorical frame, for which the rhetorical reference does not align with the policy reference. To see this, note that the game and spread victor questions can be regarded as two distinct gambles that rely on the same underlying random variable, the margin of victory, but with different thresholds for resolving the outcome --- zero and the spread respectively. As a result, their subjective probabilities yield different reference wagers as seen in Figure~\ref{fig:nflcurves}. A spread victor probability of $1/2$ corresponds to a zero wager rhetorical reference whereas a game victor probability of $1/2$ yields a negative wager as the rhetorical reference (since $1/2$ represents the belief that, counter to the oddsmakers, there is no favorite and so the spread is too high, implying an underdog bet). Thus, heuristic use of the game victor frame implies that the boundary demarcating pro and con discursive sides does not correspond to that between opposing policy sides.

The shift in reference points caused by heuristic frame substitution can cause distribution reshaping to vary by policy side. Returning to Figure~\ref{fig:riacurve}, the offset between the policy and rhetorical references signifies that the rhetorical frame displayed is an improper one. While the members of the $U$ triad are all on the same (negative) side of the policy axis, they straddle the rhetorical reference, resulting in the $U$ group being arrayed along the roughly linear part of the curve in contrast to the $F$ group on the shoulder. The $U$ group therefore is subject to much less pronounced distribution reshaping than the $F$ group: $U_2$ is nearly equidistant from both the moderate $U_1$ and the extremist $U_3$ along the rhetorical frame. The $F$ and $U$ groups are analogous to groups who choose the favorite or underdog respectively in our experiment since the rhetorical reference of the game victor frame is within the underdog wager side.

The spread betting context points to a more general condition for heuristic frame substitution and reference point shifting. They are expected whenever two different gambles, one complex and one intuitive, depend on the same scalar random variable but with different thresholds. The subjective probability of the intuitive gamble (improper frame) is then substituted for that of the complex gamble (proper frame). In investing, for example, the variable might be the return of a company's stock over a given time period. The policy gamble is whether the actual return exceeds the fixed return of a zero-risk asset such as a bond. However, group discussion may focus on the simpler intuitive gamble concerning whether the stock's price will rise or fall. The policy and intuitive gambles have different thresholds --- the fixed and zero returns respectively.

\subsection*{Rhetorically-Proximate Majority Formation} \label{sec:RPM}

This section presents the consensus formation process that accompanies distribution reshaping due to a concave rhetorical function and its implications for our experiment. The process is the same whether there is a unanimity decision requirement or group members arrive at a common policy position due purely to social influence. Figure~\ref{fig:riacurve} shows what happens if one were to suppose, at first, that consensus formation involves convergence to the mean rhetorical position of the group ($\bar{\rho}$ for the $F$ group): the corresponding policy (`a') is less than the group mean policy $x_2$ (`b') thereby signifying depolarization not polarization.  For quantitative judgments, however, the central position within a majority cluster of proximate positions has been observed to be a better predictor of consensus than the central tendency (mean or median) of the whole group \citep{DavHulAu1997,OhtMasNak2002}. Rather than convergence to the mean, we therefore root our consensus process in the formation of a rhetorically-proximate majority that emerges during the discussion --- a majority that then drives the ultimate consensus. Such emergent majorities have been found to be highly predictive of the outcomes of discrete choice experiments in what has been referred to as a ``momentum effect'' \citep{Kerr1992}.

In Figure~\ref{fig:riacurve}, the rhetorically more proximate pair is ($F_2,F_3$) and so they form a majority. Assuming they agree at some policy intermediate between them, a position to which the outlier $F_1$ then concurs, the consensus policy will be more extreme than the mean $x_2$. If we take $F_2$ and $F_3$ to have equal influence in the deliberations, the intermediate policy should result from the average of their positions. But which positions should be averaged, policy or rhetorical?  As either procedure will yield polarization, choosing between them requires an assessment of which is behaviorally more likely. Taking the average of the rhetorical positions, which then must be transformed to a policy, would represent the more rational procedure from a decision theory standpoint in that a revised subjective probability is used to determine a new optimal policy.  Doing so, however, demands that the RPM pair, whose formation relies only upon the perceived relative proximity of rhetorical positions further generate an absolute rhetorical position. In contrast, averaging their initial policy positions is a more direct procedure. An individual's initial policy is overtly numerical, unlike the rhetorical position, and so serves as the starting point that is shifted up or down during the discussion via the operation of an anchoring and adjustment heuristic in which an initial numerical judgment (anchor) forms the basis upon which subsequent adjustments are made \citep{TveKah1974}. Due to its lower cognitive demand, we therefore expect the averaging of the RPM pair policy positions to be the more typical approach when pre-discussion policy preferences have been expressed (although some groups may employ the alternative procedure in which the policy is freshly generated from the average rhetorical position of the RPM pair). Consequently, $F_2$ and $F_3$ form a majority at the average of their policy positions, $\bar{x}_{23} \equiv (x_2+x_3)/2$, which becomes the consensus policy (`c').

In summary of the RPM process above, discussion over the rhetorical frame serves to facilitate, via distribution reshaping, agreement among the more extreme members so that a majority emerges at the average of their policies. Majority influence then causes the minority to concur with the majority position. The consensus policy, and hence group mean, is more extreme than the initial mean --- the definition of group polarization. This process entails a conflation of the rhetorical frame and the policy: the rhetorical frame determines the subset of members whose \emph{policies} are aggregated to form the ultimate consensus policy. Against the dominant polarization tendency, the RPM process also allows a given group to depolarize if the moderate majority is rhetorically proximate --- an important departure from the uniform polarization prediction of standard theory.

The RPM process described above immediately yields a prediction for the effect of disagreement level on group polarization. Specifically, we consider increasing the policy distance between the two outermost members while the mean stays fixed. In Figure~\ref{fig:riacurve}, this corresponds to keeping $x_2$ fixed while moving $x_1$ and $x_3$ by equal amounts toward and away from the policy reference respectively. For the $F$ group, the RPM still consists of $F_2$ and $F_3$ but they now converge at a more extreme policy since their halfway-point is more extreme. Thus, the RPM process predicts that \emph{group polarization will increase with disagreement level}.

Turning to the effect of reference point shifting due to heuristic frame substitution, the weak distribution reshaping experienced by the $U$ group leads to the expectation of a much-reduced group polarization effect. Even if we strictly apply the RPM process regardless of how slim the difference between the rhetorical proximities of the ($U_1, U_2$) and ($U_2, U_3$) pairs ($\Delta \rho_{21}$ vs.\ $\Delta \rho_{32}$), the amount of polarization on average will still be small when uncertainty and random effects are taken into account. Small changes in rhetorical position can affect which pair is the RPM one thereby resulting in opposite direction policy shifts that cancel out on average over groups. In Figure~\ref{fig:riacurve}, the RPM position would be the average of the $U_2$ and $U_3$ policies, $\bar{x}_{23}$, which would represent a considerable shift toward the negative policy extreme compared with $x_2$. However, uncertainty and noise could readily shift the rhetorically-proximate pair to ($U_1, U_2$) so that the policy would be $\bar{x}_{12}$, representing a depolarization instead. Therefore, given the looseness with which policies correspond to rhetorical positions as well as stochastic effects in the discussion itself, the final policy over many groups will average out to nearly $x_2$ yielding little systematic polarization. In contrast, the effect of the offset rhetorical reference upon the $F$ groups is to shift them further toward the shoulder of the rhetorical function, thereby increasing the distribution reshaping effect. Synthesizing, we arrive at the following prediction: \emph{When heuristic frame substitution results in an offset between the rhetorical and policy references, group polarization will be enhanced on the policy side that does not contain the rhetorical reference and suppressed on the side that does}. For the experiment, this implies that favorite groups will exhibit greater group polarization than underdog groups.

In addition, our consideration of network structure focuses on the situation where the network has a central member with more connections than other members in the group. To the extent that this affords the central node a communication advantage, the RPM position and, hence, the consensus policy will shift toward the central node's initial policy position (if belonging to the RPM). In our experiment, the center node in the chain is also taken to be in the middle of the policy distribution as such an alignment would be likely to arise under a tie formation process guided by homophily.  For both chain and complete networks, distribution reshaping favors majority formation between the intermediate and extreme policy nodes. However, the chain center node's communication advantage implies that the chain consensus policy will be less extreme than for complete networks. Consequently, we predict that \emph{complete networks will exhibit greater group polarization than chains}.

\subsection*{Formal RPM Model}

This section formalizes the RPM process described above. We consider a triad whose members have initial policy positions $(x_1,x_2,x_3)$, where $x_2$ is intermediate between $x_1$ and $x_3$, and rhetorical positions $\rho_j\equiv\rho(x_j)$.  Denoting the difference $\Delta\rho_{ij}=\rho_j-\rho_i$, the final group policy $x_f$ in the RPM model is given by the weighted average of the pair with the smaller rhetorical distance:
\begin{equation}
x_f=\left\{
\begin{array}{cc}
     (\nu_{12}x_2+\nu_{21}x_1)/(\nu_{12}+\nu_{21}), & |\Delta\rho_{12}| \leq |\Delta\rho_{32}| \\
     (\nu_{32}x_2+\nu_{23}x_3)/(\nu_{32}+\nu_{23}), & |\Delta\rho_{12}| > |\Delta\rho_{32}|,
\end{array}
     \right.
\label{eq:rpm}
\end{equation}
where the communication weights $\nu_{ij}$ account for the communication rate from $j$ to $i$. Note that only distances between rhetorical positions appear here, not absolute rhetorical positions.

We treat the communication weights using simple topological considerations. For a complete network, we expect that, on average, the communication rates will be the same for all nodes, so we set $\nu_{ij}=1/2$ for all three dyads.  For the chain, if the sequence in which nodes send messages follows the chain path and the center node (node 2) predominantly opts to send its messages simultaneously to both outer nodes (rather than separately), then we expect node 2 to have about twice the communication rate with each of nodes 1 and 3. We therefore set $\nu_{12}=\nu_{32}=1$ and $\nu_{21}=\nu_{23}=1/2$. These communication rate expectations are approximately borne out in our experiment.\endnote{See online Table~B.1} The consensus policy in the complete network is given by the straight average of the policies of the rhetorically-proximate pair. In the chain, however, the RPM position is skewed toward the center node given that it has twice the weight of the extreme node. The consensus policy for the chain is therefore less extreme than for the complete network ($x_f=(2/3)x_2+(1/3)x_3$ vs.\ $x_f=(1/2)x_2+(1/2)x_3$) as argued above.

Simulations show the ability of the RPM model to predict the hypothesized behaviors for disagreement level, reference point shifting, and network structure in the presence of variation in the intermediate policy $x_2$.\endnote{See online Appendix~A.2} A straightforward way of extending the RPM model to larger networks is to define the rhetorically-proximate majority as the set of nodes comprising the majority that spans the minimum range over the rhetorical frame. The final consensus policy is then given by the weighted average policy within this set; each node's policy is weighted by the sum of its outgoing communication weights to the other nodes in the set divided by the sum of all the communication weights among set members.

\section*{Methods}

Triads of NFL fans engaged in online discussions concerning how much to wager on the outcome of a weekly NFL game (Figure~\ref{fig:setup}).\endnote{This human subjects research was approved by the University of Washington Institutional Review Board and the Defense Threat Reduction Agency Research Oversight Board.} The wager was with respect to the point spread. To provide real stakes to the task, discussion participants were given a \$7 bonus, to be donated to a specified charity, of which they could wager all, some, or none. Our use of football games was motivated by the desire to give participants a task in which they could draw on their natural knowledge base to forecast the outcome of a real-world event. While other experiments have employed betting contexts (\citealt[608--609]{MyersLamm1976}; \citealt[1142--43]{Isenberg1986}), groups did not take opposite sides of the same bet as with our policy side condition. Another distinctive feature of our experiment is control of the policy distribution (both mean and variance) of groups on both sides, unlike past experiments which simply endeavored to compose homogeneous groups. This tight control enables us to study the effects of policy side in an evenhanded way as well as disagreement level. The use of (closely) symmetric policy distributions within each group eliminates \emph{policy-based} majority influence as a cause of polarization.

\emph{Recruitment.} Recruitment took place via Amazon Mechanical Turk. Informed consent was obtained from all study participants. An initial survey included demographic questions, an NFL knowledge test, and questions on risk tolerance, argumentation, and group discussion. The knowledge screening consisted of 19 questions that asked respondents to correctly pair team locations and names and match current season players with teams. Those scoring 80 percent and above were eligible to take part in the full experiment. Demographically, most discussion participants were young white males.\endnote{See online Table~B.2.} To generate sufficient sample size, subjects were allowed to participate in multiple weeks.\endnote{Out of 240 unique participants in the consensus groups, 126 were repeat ones. As the task involved an upcoming real-world event with an unknown outcome, not an artificial game that could be learned, repeat participants had no forecasting advantage over first-timers. Repeat participants showed no greater tendency toward favorites or underdogs than first-timers nor toward making different wagers (see online Table~B.3). See Note~\ref{ref:tests} however.}

\emph{Weekly pre-survey.} In a weekly pre-survey, given 2-3 days prior to the discussion, subjects were asked to: (1) make a choice as to whether or not the favorite's margin of victory would exceed the spread; and (2) make a wager on their team choice ranging from \$0 to \$7 in whole dollar increments.\endnote{See online Appendix~C.1. A \$0 wager person may lean toward one team but not confidently enough to wager (see prospect theory $w(p)$ curve in online Figure~A2).}

\emph{Group assignment.} A subset of the pre-survey subjects was used to populate discussion groups with respect to three dichotomous variables: (1) Policy side conditions were favorite and underdog in which all members made the same team choice. (2) Disagreement level, the difference between the highest and lowest wagers in the group, was either high  or low  with the possible wager distributions shown in Figure~\ref{fig:setup}. High disagreement groups had a \$7 difference and low could be \$3, \$4, or \$5. (3) Network structure was either complete or chain, in which the moderate ($w_1$) and extreme ($w_3$) wager members were connected only to the intermediate-wager ($w_2$) member.\endnote{Not all 8 conditions were present in every game due to the need to mitigate substantial attrition caused by non-attendance of invited participants.} Complete network members sent and received messages from all others by default. The center node in the chain had the option of sending messages to only one or both of the other members (in practice the latter option was much more frequently used).

\emph{Discussion procedure.} Discussions were unstructured except for the requirement for each individual to make a final wager at the end.  Subjects could change their wager but not their team choice. Group discussion took place online using a text-based chat system.  Upon logging in, participants were presented with a series of four brief instruction screens that: (1) informed them they would be engaged in a discussion concerning the upcoming game in which they could earn money for charity; (2) described the structure of their communication network including a pictorial representation showing which node they occupied; (3) told them that all the members of the group had chosen the same team and presented the wagering options; and (4) gave instructions on how to use the chat interface and when and how to make final wagers.\endnote{See online Appendix~C.2} Participants were required to discuss the topic for a minimum of 20 minutes before they could call for a vote in which they chose their wagers, which were displayed on the chat interface. The discussion could continue up to 30 minutes. No requirement for consensus was imposed and subjects could make different final wagers.\endnote{In one game (Week 19), a consensus decision rule was mistakenly set in the experimental software so that it required groups to vote again if they did not reach consensus on the first vote. Participants were unaware of this, however, as the instructions were the same as in the other games. Consequently, we used the first vote of the groups that did not have consensus first votes (only 2 of 29 groups).} The topic, rules, and wagering options were accessible during the discussion through a help button. After registering their choices participants were directed to a short post-survey that asked several follow-up questions including their assessment of group influence and their satisfaction with the outcome.

\emph{Group data.} We collected data for six NFL games taking place between December 2014 and February 2015. We excluded 24 groups whose transcripts showed no discussion, had users drop out prematurely, or did not make final wagers. This resulted in 198 valid groups. The distribution of outcomes was: 159 consensus groups; 30 where only two members made the same wager; and 9 where all three wagers were different.\endnote{See online Table~B.4 for games, online Table~B.5 for group descriptive statistics, and online Appendix~D for group wagers.}

\emph{Compensation and donations.} Participants were paid a flat amount for participation in the discussion: \$8 during regular season play and \$16 during the playoffs. Participants were informed that the \$7 charity bonus they received plus (minus) their winning (losing) wager amount would be donated to the Wounded Warrior Project, a charity set up to aid wounded US military veterans.\endnote{Winnings went to charity rather than participants to avoid potentially adverse public appearances of sports gambling. A total of \$6756.50 was donated.} Participants' performance in raising charitable donations was tracked using an online leaderboard.\endnote{The increase in playoffs pay and the leaderboard were intended to motivate subjects to take the task seriously and continue participating. There was no requirement to check the leaderboard.}

\section*{Results}

\subsection*{Tests of Qualitative Hypotheses}

We analyze only the consensus groups, which was by far the most common outcome (80 percent). This is consistent with our theory and most past experiments. To quantify group polarization, we calculate the \emph{group shift} defined as the change in the mean wager of the group after discussion: $\delta=w_f-\bar{w}_0$ where $\bar{w}_0$ is the initial mean wager and $w_f$ is the post-discussion mean (consensus) wager.  A positive $\delta$ indicates an increase in the mean wager and hence group polarization, more specifically, a risky shift. A negative $\delta$ indicates a cautious shift.

Table~\ref{tab:shifts} shows the mean group shift $\bar{\delta}$ for specified variable conditions along with the difference between the $\bar{\delta}$ values for the paired conditions. Statistical analysis was performed using Student's \emph{t}-test for the tests of $\bar{\delta}$ relative to a null shift and Welch's \emph{t}-test for the difference $\Delta\bar{\delta}$ between two compared conditions. Two-tailed testing was employed and the null hypothesis was rejected for $p<.05$. For policy side, the favorite groups exhibit a risky shift of \$1.45 which is highly significant compared with a null shift ($p=10^{-11}$) whereas the underdogs show an insignificant positive shift of \$.19. Naturally, the difference between their shifts ($\Delta\bar{\delta}=\$1.26$) is also  significant ($p=.00007$). Consequently, we conclude that the favorite groups exhibit a risky shift and the underdog groups do not. Regarding disagreement level and network structure, for favorite groups: high disagreement shows a significantly greater shift than low ($\Delta\bar{\delta}=\$.92,p=.012$) and the complete network shows a significantly greater shift than the chain ($\Delta\bar{\delta}=\$1.01,p=.009$). Each favorite condition on its own displays a shift that is significantly greater than null whereas none of the underdog disagreement or network conditions do, consistent with the overall absence of an underdog risky shift.

\begin{table}[p]
\vspace{1in}
\small
\begin{tabular*}{\hsize}{@{\extracolsep{\fill}}lrllllll}
\textbf{Condition}  &
\multicolumn{1}{c}{N} &
\multicolumn{1}{c}{$\bar{\delta}$ (\$)} &
\multicolumn{1}{c}{SE (\$)}  &
\multicolumn{1}{c}{$p(\bar{\delta})$} &
\multicolumn{1}{c}{$\Delta\bar{\delta}$ (\$)} &
\multicolumn{1}{c}{$p(\Delta\bar{\delta})$}  &
\multicolumn{1}{c}{\emph{t}(\emph{df})}
\\ \hline\hline

Favorite  & 103 & 1.45*** & .19 & $1\times 10^{-11}$ & \raisebox{-1.5ex}{1.26***} & \raisebox{-1.5ex}{.00007} & \raisebox{-1.5ex}{4.12 (119.4)} \\
Underdog  & 56  & .19 & .24 & .43   &       &              & \\ \hline
Fav./High  & 59 & 1.84*** & .27 & $4\times 10^{-9}$ & \raisebox{-1.5ex}{.92*} & \raisebox{-1.5ex}{.012} & \raisebox{-1.5ex}{2.55 (100.7)}\\
Fav./Low  & 44  & .92*** & .24 &  .0004 &         &              & \\ \hline
Fav./Comp.  & 37 & 2.10*** & .30 & $3\times 10^{-8}$ & \raisebox{-1.5ex}{1.01**} & \raisebox{-1.5ex}{.009} & \raisebox{-1.5ex}{2.69 (77.5)} \\
Fav./Chain  & 66  & 1.09*** & .23 &   .00002  &      &                & \\ \hline
Und./High  & 22 & .27 & .54 & .62 & \raisebox{-1.5ex}{.14} & \raisebox{-1.5ex}{.82} & \raisebox{-1.5ex}{.24 (26.6)}\\
Und./Low  & 34  & .14 & .20 &   .49 &        &              & \\ \hline
Und./Comp.  & 31 & .14 & .32 & .67 & \raisebox{-1.5ex}{$-$.11} & \raisebox{-1.5ex}{.82} & \raisebox{-1.5ex}{$-$.23 (51.0)}\\
Und./Chain  & 25  & .25 & .37 &    .50 &     &                & \\ \hline
\end{tabular*}

* $p<.05$, ** $p<.01$, *** $p<.001$

\caption{Mean group shifts for specified conditions using the 159 groups with consensus outcomes. N is the number of groups; $\bar{\delta}$ is the mean of $\delta=w_f-\bar{w}_0$ taken over the N groups; SE is its standard error and $p(\bar{\delta})$ is its statistical significance level as compared to a null shift; $\Delta\bar{\delta}$ is the difference between the $\bar{\delta}$ values of the top and bottom variables in the comparison pair and $p(\Delta\bar{\delta})$ is its statistical significance level; $t$ and \emph{df} are the \emph{t}-test t-score and degrees of freedom respectively.} \label{tab:shifts}

\end{table}

These results confirm all three of our hypotheses above.\endnote{\label{ref:tests} To check for the possible effects of non-normal distributions, nonparametric testing was conducted (see online Table~B.6). Very similar results to Table~\ref{tab:shifts} were obtained. A regression analysis shows a significant tendency for groups with repeat participants to decrease the group shift as compared with groups with no repeat participants, but our three manipulated variables remain significant (see online Table~B.7).} Favorite groups polarize more than underdog groups: heuristic substitution of the improper game victor frame instead of the proper spread victor frame shifts the rhetorical reference to lie within the underdog policy side, thereby subjecting underdog groups to weaker distribution reshaping. High disagreement groups polarize more than low: the extremist has a higher wager in the former so that the policy of the RPM forms further toward the extreme. Complete networks polarize more than chains: the communication rate advantage of the chain center node causes the RPM policy to form closer to the center node's intermediate wager than the even split between intermediate and extreme wagers in complete networks. As discussed above, these results are not readily explained by standard polarization theory. The differential polarization by policy side, in particular, is directly counter to standard theory's expectations.

\subsection*{RPM Model Quantitative Performance} \label{sec:simresults}

In this section, we test the quantitative performance of the RPM model. To do so, requires using a specific form for the rhetorical function. If our subjects were professional gamblers, who presumably would not engage in heuristic substitution, we could use the subjective probability of a favorite spread victory $p(w)$ (see Note~\ref{note:pw}). Its S-shape in Figure~\ref{fig:nflcurves} (concavity with respect to increasing wager magnitudes) implies that even groups of expert gamblers would exhibit polarization induced by distribution reshaping. Yet, as a proper rhetorical frame, differential polarization by policy side would not be expected.

The expectation of heuristic frame substitution among our subjects, however, calls for using the subjective probability of a favorite game victory as the frame. NFL fans have been found to have an intuitive understanding of the concave relationship between the favorite game victory probability and the expected margin of victory --- that the probability flattens out as the margin increases \citep[422]{SimmNel2006}. In the derivation of the rhetorical function, $\rho(w)$, this fact motivates taking the victory margin to be a normally distributed random variable with standard deviation $\sigma$ and whose mean, $s$, corresponds to an individual's estimate for the victory margin. The spread, $s_0$, is the estimate of the oddsmakers. Figure~\ref{fig:nflcurves} shows that the derived form of $\rho(w)$ is S-shaped.\endnote{\label{note:rhow}$\rho(w) = \frac{1}{2} \mathrm{erfc} \left\{ \mathrm{erfc}^{-1}\left(2 p(w)\right) - \frac{s_0}{\sigma \sqrt{2}} \right\}$ where $\mathrm{erfc}(x)=\frac{2}{\sqrt{\pi}}\int_{x}^{\infty} e^{-v^2}dv$. See online Appendix~A.3.}

We use $\rho(w)$ in our simulations assuming identical risk aversion $\alpha$ and standard deviation $\sigma$.\endnote{Use of the identical risk aversion assumption is supported by the poor correlation ($-.02$) between initial wagers and risk tolerance as assessed from the initial screening survey.} Rather than using $\alpha$ directly, we define the more intuitive parameter $p_{max}$ which corresponds to the subjective probability $p(w_{max})$ for which an individual will make the maximum wager, $w_{max}=\$7$.\endnote{$\alpha = \frac{2p_{max}-1}{8 w_{max} p_{max} (1-p_{max})}$.}  While $p_{max}$ is a free parameter that must be estimated using the data, we fix $\sigma=12.8$, which is the empirical standard deviation for the margin of victory in NFL games \citep{Sapra2008}. $s_0$ is set to the spread for the game of concern. Each group is simulated using its initial wagers $w_1, w_2, w_3$ to find the consensus policy $w_f$ as given by Equation~\ref{eq:rpm} (with $w_i$ replacing $x_i$). As subjects could only make whole dollar wagers, the RPM model is supplemented with a normative rounding scheme: a norm toward risk-taking is assumed so that final simulated wagers are rounded upward to the nearest dollar for both favorite and underdog groups.

\begin{figure}[p]
\vspace{0.5in}
\begin{minipage}{0.5\textwidth}
\centering
\includegraphics[scale=0.5]{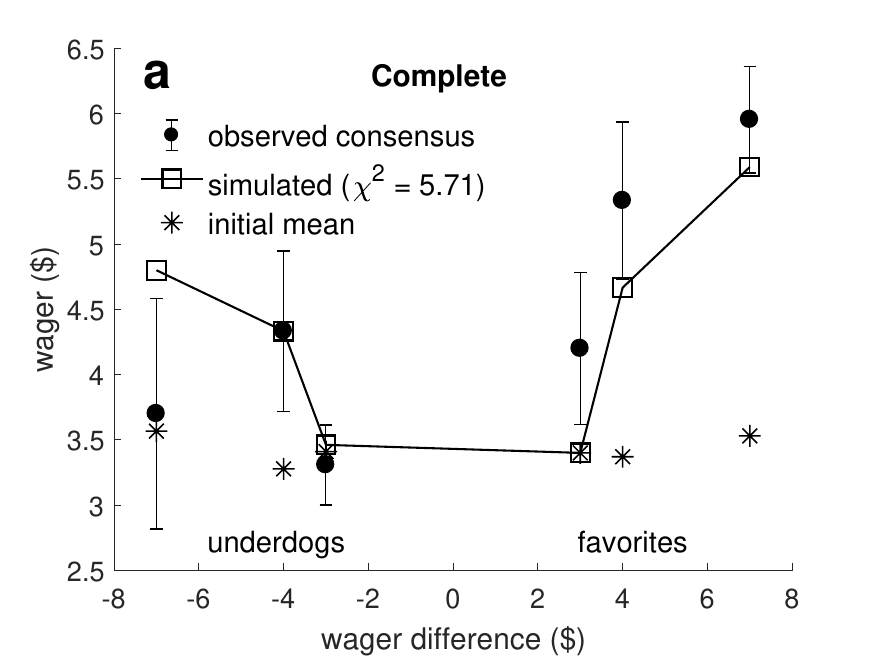}
\end{minipage}
\begin{minipage}{0.5\textwidth}
\centering
\includegraphics[scale=0.5]{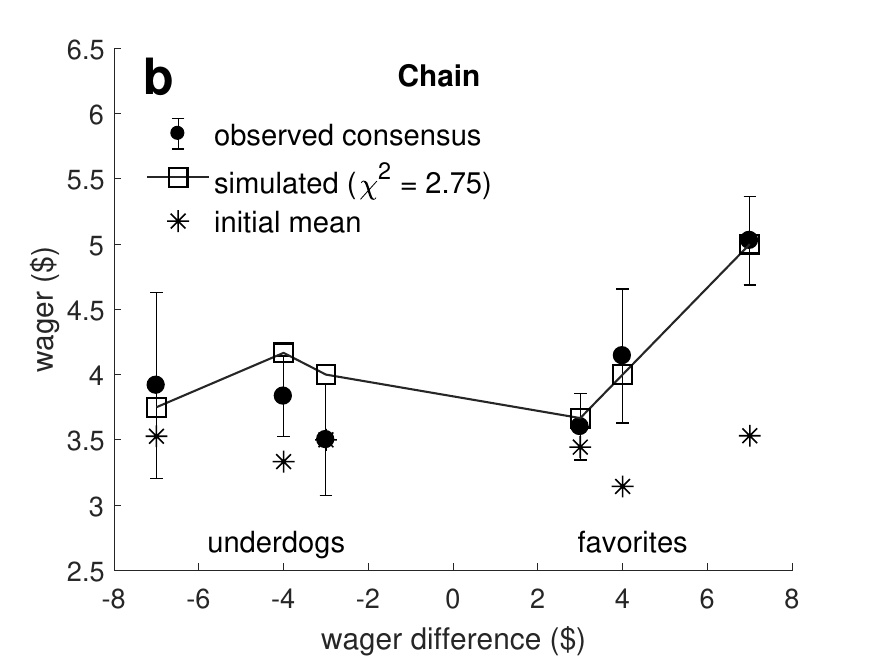}
\end{minipage}
\caption{Observed and simulated mean consensus wagers as function of initial wager difference, $w_3-w_1$. (\textbf{a}) Complete network. (\textbf{b}) Chain network.  Favorite groups shown on the positive $x$ axis, underdogs on the negative. Mean consensus wager is average of $|w_f|$ over groups at each difference value. Also shown is average of the group mean initial wager $\bar{w}_0$.  Experimental data shown as circles. Error bars are standard errors. Sample sizes from left to right: Complete: 10, 6, 13, 5, 9, 22; Chain: 12, 6, 6, 15, 14, 37 (no \$5 disagreement groups were used as there were only four total).  $p_{max}=.71$. $\chi^2$ value is the sum of the squared errors between the simulated and the experimental values normalized by the standard error at each data point.}
\label{fig:expplot}
\end{figure}

Figure~\ref{fig:expplot} compares the observed and simulated consensus wager as a function of the initial wager difference, $w_3-w_1$, averaged over the groups at each difference. Polarization is (loosely) indicated if the observed error interval is higher than the initial mean wager. Qualitatively, the simulations display the hypothesized behaviors of greater polarization for favorites, high disagreement, and complete networks. The value of the free parameter $p_{max}$ is selected so as to minimize the $\chi^2$ error summed over both networks. A one-parameter $\chi^2$ goodness-of-fit test --- which takes as its null hypothesis that the model is correct --- yields a probability $Q=.67$ that $\chi^2$ could have exceeded the observed sum of 8.46 by chance. With a conservative threshold of $Q < .2$ (as has been used previously: \citealt{DavHulAu1997}; \citealt{OhtMasNak2002}) for rejecting the null hypothesis, the test finds the RPM model to be consistent with the data.

We have also tested alternative models, all of which fail: (1) the median wager ($Q=2\times10^{-10}$); (2) a policy-based proximate majority rule based directly on the wager ($Q=8\times10^{-4}$); (3) an intuitive confidence-based policy averaging with wagers weighted by the game victor subjective probability ($Q=.026$).\endnote{See online Appendix~A.4.}

We note that the RPM model systematically undershoots the data for the complete network favorites. We have no definitive explanation for this undershoot but offer the following speculation, consistent with the pivotal role of rhetorically-proximate majority formation in our theory. The direct interaction between the moderate ($w_1$) and extreme ($w_3$) members afforded by the complete network exposes them to stronger challenges from each other than in the chain. If formation of the RPM pair is characterized as a contest between them over which way the intermediate member ($w_2$) will shift, the successful rebuttal of such challenges may make the ``winner's'' case appear more convincing. Consequently, the intermediate member may shift further toward the winner than the even split assumed in the RPM model. For the favorite groups, the RPM pair more frequently forms on the extreme side, and so not accounting for this presumed greater shift results in a systematic undershoot. For the underdog groups, however, the RPM pair is roughly equally likely to form on either the moderate or extreme sides and so the neglected greater shift of the intermediate member occurs as often toward lower wagers as toward higher ones, yielding no net effect on the observed mean. This explanation assumes a more dynamic and interactive view of argumentation than the recitation of preset, static arguments entailed by persuasive arguments theory \citep{SeiMey2007}. Another limitation of the RPM model is that it cannot predict policies outside the range spanned by the minimum and maximum policies in the group, which occurs in 14.1 percent of the low disagreement groups. Mathematically, one approach to accommodating such cases is by generating the policy stochastically from the RPM pair instead of deterministically.

\section*{Discussion}

We have proposed a frame-induced mechanism for group polarization that is supported by our experimental results. This process can operate concurrently with the informational and normative mechanisms of standard polarization theory. Highlighting key differences with standard theory, however, can help illuminate the nature of our theory. First, the way an issue is discussed, the rhetorical frame, is central to our mechanism, not the exchange of new information or an issue-salient norm that operates even absent discussion. Persuasive speech is therefore essential to our mechanism and consensus can arise through a mix of informational and normative influence processes (beyond the specific standard theory processes). In a particular group, for example, the majority may form due to informational influence while the minority conforms to the majority due to normative influence. Second, majority influence and consensus pressure are integral elements of our mechanism thereby embedding our account within the broader attitude change landscape rather than standing apart from it. This incorporation of disagreement reduction processes enables us to account for disagreement level and also allows for the possibility of particular groups depolarizing despite an overall tendency toward polarization. The reliance on majority formation, however, implies that our mechanism, unlike standard theory, cannot address polarization in dyads.

A third difference is that the focus on the rhetorical frame allows us to take more explicit account of the reference point, opening up the possibility that the policy and rhetorical references need not align as occurs with the substitution of an improper rhetorical frame. Fourth, while the processes in the frame-induced and standard polarization theories all transpire at the individual group level, our theory can, via the RPM model, also make predictions at that level. The RPM model can predict gradations in behavior for specific preference distributions, enabling the quantitative testing of modeled outcomes versus the data. Furthermore, in standard theory the reference point is tantamount to a chasm: polarization is clearly predicted for homogeneous groups, but application to heterogeneous groups, which span both sides, is ambiguous. In the RPM model, the rhetorical reference simply marks the center of the linear portion of the S-shaped rhetorical function, neither imposing uniform polarization on homogeneous groups nor preventing the modeling of heterogeneous groups. This reflects that the RPM model can serve as a broader model of opinion change, not just polarization.

Another important aspect of our work is that the rhetorical function for subjective probability is grounded in the theory of decision making under risk and uncertainty. Its concavity even for a proper rhetorical frame suggests that policy experts, presumably less prone to heuristic substitution, will also exhibit polarization. This concavity arises from risk aversion, a cornerstone of decision theory, and so is more generally applicable than the experimental context. That distribution reshaping induces a risky shift then implies that group discussion tempers the risk aversion of individuals.  Although recent economics literature has found mixed results regarding whether groups are less risk averse than individuals \citep{Sutter2007,BakLauWil2008}, these experiments have involved lotteries with known probabilities, whereas probability uncertainty is intrinsic to our experiment and theory and to attitude change more broadly. We also provide specific guidance regarding when to expect reference shifting due to heuristic frame substitution: when the policy gamble (e.g., the spread victor) and the heuristic gamble (the game victor) both depend on the same random variable (the victory margin) but with different thresholds.

Beyond quantitative policies, we expect the frame-induced polarization mechanism to apply more generally to attitudes. An individual's attitude toward a target integrates their evaluations over various attributes. Each evaluation depends upon their belief (subjective probability) that the target possesses that attribute and the value they place on it. Accordingly, if the discussion focuses upon a rhetorical frame concerning the probability that the target possesses a particular attribute (rather than a more comprehensive consideration of multiple attributes), then distribution reshaping is expected. Similarly, the Likert scale measures an individual's attitude by aggregating their reported level of agreement with a number of items, each of which expresses a positive or negative opinion toward the target. An individual's level of agreement with an item is (ideally) an S-shaped function of their overall attitude \citep{RobLauWed1999}, and so concave with respect to increasing favorable or unfavorable extremity. Consequently, a discussion centered on one item would bias majority formation toward the extreme, implying polarization.

Although polarization has been our concern here, a convex rhetorical function will expand distances at the extreme relative to distances closer to the center thereby favoring the formation of moderate majorities. Thus, convex rhetorical functions would be expected to produce a tendency for groups to depolarize. As it is possible that an issue could have both concave and convex rhetorical frames, the same issue could result in polarization or depolarization contingent upon which frame is used.

Though we believe that the frame-induced mechanism is theoretically compelling and corroborated by our data, further experiments can more conclusively establish its causal process. Directly estimating the rhetorical function by eliciting subjective probabilities from subjects in addition to wagers would be one approach (however, the capability we demonstrated of inferring subjective probabilities directly from the central inputs and outputs of the discussion task itself, the wagers, is more applicable to natural settings). Another would be manipulation of the rhetorical frame so that some groups discuss proper frames and others improper ones. Additionally, the observation of polarization and depolarization for the same issue due to concave versus convex frame manipulation would be strong evidence in support of the present theory.

\section*{Supplemental Material}
Additional supporting information may be found at: \\ {\scriptsize{\url{https://journals.sagepub.com/doi/suppl/10.1177/0190272518778784}}}.

\section*{Acknowledgments}
We thank the three anonymous reviewers and the editors for their insightful comments and suggestions. We also thank Michael Farrell, Leonie Huddy, Joong Kim, Caren Marzban, and participants of various conferences at which earlier versions of this work were presented for helpful discussions and remarks.

\section*{Funding}
This work was supported by the Defense Threat Reduction Agency and the Office of Naval Research under grants HDTRA1-10-1-0075 and N00014-15-1-2549.

\theendnotes

\section*{Bios}

\textbf{Michael Gabbay} (PhD, University of Chicago) is a senior principal physicist at the Applied Physics Laboratory of the University of Washington. His current research focuses on the application of network science and complex systems theory to the modeling and analysis of social influence processes, group decision making, and factional dynamics within political leadership networks and militant movements.

\textbf{Zane Kelly} (PhD, University of Colorado-Boulder) is currently a data scientist at University of Washington-IT, in the Academic Experience Design and Delivery team where he helps develop technology tools for students, faculty, and staff. Previously, he was a postdoctoral researcher at the Applied Physics Laboratory of the University of Washington.  His research interests include social network analysis and political economy.

\textbf{Justin Reedy} (PhD, University of Washington) is an assistant professor in the Department of Communication and research associate in the Center for Risk and Crisis Management at the University of Oklahoma. His research focuses on how groups of people make political and civic decisions in face-to-face and online settings, how people and policy makers can come together to deliberate on risk-related policy issues, and how group communication theories can help us better understand terrorist and extremist groups.

\textbf{John Gastil} (PhD, University of Wisconsin-Madison) is a professor in the Department of Communication Arts and Sciences and Political Science at the Pennsylvania State University, where he is a senior scholar at the McCourtney Institute for Democracy. Gastil's research focuses on the theory and practice of deliberative democracy, particularly as it relates to how people make decisions in small groups on matters of public concern. His most recent books include \emph{The Jury and Democracy} and a second edition of \emph{Democracy in Small Groups}.

\end{document}